\documentclass[journal,twocolumn]{IEEEtran}
\usepackage{graphicx,times,amsmath,amssymb,cite,subfigure,stfloats,booktabs, url,multirow,afterpage}
\usepackage{graphicx}
\usepackage{caption}
\usepackage{array}
\usepackage{booktabs}
\usepackage{commath,amsfonts,algorithm,algorithmic}
\usepackage{cite,multirow,bm,multicol,booktabs,threeparttable,extpfeil}

\allowdisplaybreaks

\begin{document}
\title{CLEAN-MI: A Scalable and Efficient Pipeline for Constructing High-Quality Neurodata in Motor Imagery Paradigm}

\author{Dingkun~Liu, Zhu~Chen, and Dongrui~Wu,~\IEEEmembership{Fellow,~IEEE}
\thanks{D.~Liu, Z.~Chen and D.~Wu are with the Ministry of Education Key Laboratory of Image Processing and Intelligent Control, School of Artificial Intelligence and Automation, Huazhong University of Science and Technology, Wuhan 430074, China.}
\thanks{D.~Liu and D.~Wu are also with Zhongguancun Academy, Beijing, 100080 China.}
\thanks{Corresponding Authors: Dongrui Wu (drwu09@gmail.com).}
}

\maketitle

\begin{abstract}

The construction of large-scale, high-quality datasets is a fundamental prerequisite for developing robust and generalizable foundation models in motor imagery (MI)-based brain–computer interfaces (BCIs). However, EEG signals collected from different subjects and devices are often plagued by low signal-to-noise ratio, heterogeneity in electrode configurations, and substantial inter-subject variability, posing significant challenges for effective model training. In this paper, we propose CLEAN-MI, a scalable and systematic data construction pipeline for constructing large-scale, efficient, and accurate neurodata in the MI paradigm. CLEAN-MI integrates frequency band filtering, channel template selection, subject screening, and marginal distribution alignment to systematically filter out irrelevant or low-quality data and standardize multi-source EEG datasets. We demonstrate the effectiveness of CLEAN-MI on multiple public MI datasets, achieving consistent improvements in data quality and classification performance.

\end{abstract}


\section{Introduction}

A brain-computer interface (BCI) serves as a direct communication pathway between the human or animal brain and an external device~\cite{nicolas2012brain}. There are generally three paradigms of BCIs: motor imagery (MI), steady-state visual evoked potentials (SSVEP), event-related potential (ERP). The MI paradigm, widely studied for its significant role in medical applications such as stroke rehabilitation, is the most extensively researched and applied BCI paradigm.

\begin{figure}[htbp]         \centering
\includegraphics[width=\linewidth,clip]{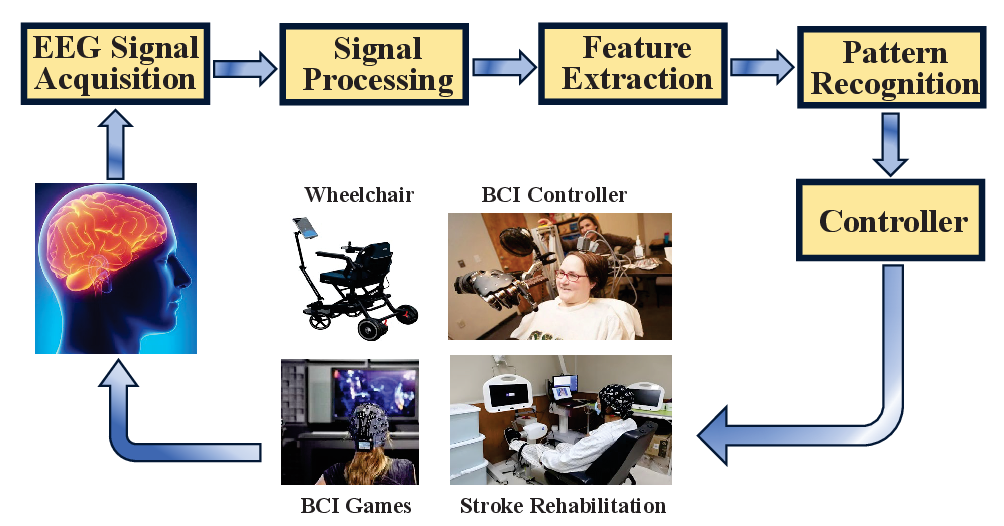}
\caption{A closed-loop MI-based BCI system.} \label{fig:MI_pipeline}
\end{figure}

The pipeline of a closed-loop MI-based BCI system is shown in Fig.~\ref{fig:MI_pipeline}. It consists of the following main components:

1. \textbf{EEG signal acquisition}: EEG signals are acquired using a headset with conductive paste applied to ensure good contact with the scalp. The subject then performs motor imagery tasks based on on-screen cues, with EEG signals recorded during the task.

2. \textbf{Signal processing}. EEG signals in MI paradigm are acquired from the subject's scalp, which is distant from the cortical sources of brain activity. As a result, these signals often exhibit a low signal-to-noise ratio (SNR) and include components from multiple frequency bands. The alpha ($\alpha$) and beta ($\beta$) rhythms, which are significant to MI, are typically selected by applying bandpass filtering in the 8-30 Hz range. Additionally, to address inter-subject variability in EEG signals, alignment techniques, such as Euclidean alignment (EA)~\cite{he2019transfer}, are commonly used to map the signals from different subjects into a consistent spatial domain. Spatial filtering methods, including common spatial pattern (CSP), are often employed to enhance the discriminability of MI tasks by extracting spatial features that improve classification performance. However, CSP is primarily effective in the MI paradigm and may not be suitable for other paradigms, such as P300, where methods like xDAWN are more commonly used for spatial filtering.

3. \textbf{Feature extraction}. Feature extraction involves identifying relevant features from the processed EEG signals, which can be categorized into time-domain, frequency-domain, and time-frequency-domain features. In addition to traditional machine learning techniques like linear discriminant analysis (LDA), AdaBoost, and support vector machines (SVM), deep learning approaches can also be utilized to automatically extract features from raw EEG data. These approaches have shown promise in learning more complex and higher-level representations of the EEG signals for improved classification accuracy.

4. \textbf{Classification}. After feature extraction, EEG features are used for pattern recognition, typically through linear projection methods like multi-layer perceptron (MLP).

5. \textbf{Controller}: The controller issues commands to external devices, such as a wheelchair or robotic arm, based on the decoded EEG signals and classification results.

Following the aforementioned pipeline, numerous specialized models have been designed to address specific EEG tasks. With recent rapid advancements in large-scale pretraining techniques, constructing general-purpose EEG foundation models capable of adapting efficiently to diverse downstream tasks has become both feasible and increasingly desirable. Preliminary studies on EEG foundation models have demonstrated promising outcomes by pretraining on extensive multi-paradigm EEG datasets and performing subsequent fine-tuning on downstream tasks. However, such approaches have not sufficiently addressed two critical issues: (1) substantial differences exist among EEG paradigms regarding data acquisition methodologies, active cortical regions, underlying neurological principles, and relevant frequency bands; and (2) practical deployment scenarios typically enable the identification of the required paradigm before the downstream task data become available. Thus, building paradigm-specific foundation models emerges as a more effective and practically justified research direction.

In this paper, we specifically focus on developing an efficient data construction pipeline tailored for general-purpose foundation models within the motor imagery (MI) paradigm. MI foundation models aim to leverage extensive MI datasets during the pretraining stage and achieve rapid adaptation on downstream MI tasks with minimal calibration data. However, significant challenges arise due to variations in the number and positioning of EEG channels across different MI datasets. Therefore, an effective and systematic approach to channel selection is essential. Moreover, non-invasive EEG data collection typically yields signals characterized by low signal-to-noise ratio. Additional variability in data quality often arises from participant inattentiveness and experimental noise, further complicating model training. Hence, selecting high-quality subjects and ensuring data integrity are crucial for constructing large-scale and high-quality EEG datasets to facilitate robust MI foundation model training.

To address the challenges of constructing large-scale, high-quality EEG datasets for motor imagery (MI) foundation models, we propose a scalable and efficient pipeline for Constructing Large-scale Efficient and Accurate Neurodata for MI (CLEAN-MI). This pipeline is designed to handle channel inconsistency, data noise, and subject variability, serving as a robust data foundation for pretraining general-purpose MI models.

The main contributions of this paper can be summarized as follows:

\begin{itemize}
\item To the best of our knowledge, this is the first work to propose a systematic pipeline for constructing large-scale, high-quality EEG datasets specifically for MI foundation models.

\item We introduce a well-defined MI channel template to identify EEG channels closely associated with MI tasks, thereby enhancing signal relevance and computational efficiency.

\item We propose an effective subject selection module, enabling the exclusion of low-performing subjects whose data may degrade the foundation model performance.
\end{itemize}

The remainder of this paper is organized as follows. Section~\ref{sect:related} introduces related work. Section~\ref{sect:method} proposes CLEAN-MI. Section~\ref{sect:Datasets} provides an overview of MI public datasets. Section~\ref{sect:exp} presents the experiment results. Section~\ref{sect:future} discusses the future work. Finally, Section~\ref{sect:future} draws conclusions.

\section{Related Work} \label{sect:related}

This section introduces related works on heterogeneous transfer learning and EEG foundation models.

\subsection{Heterogeneous Transfer Learning}

Recently, a few cross-dataset transfer learning approaches have been explored in EEG-based BCIs. Wu \emph{et al.}~\cite{drwuTNSRE2016} proposed active weighted adaptation regularization, which integrates domain adaptation and active learning, for cross-headset transfer. Xu \textit{et al.}~\cite{xu2021enhancing} combined alignment and adaptive batch normalization in neural networks, also integrating manifold embedded knowledge transfer~\cite{Zhang2020} to improve generalization. Xie \textit{et al.}~\cite{xie2023cross} proposed a pretraining-based cross-dataset transfer learning approach for MI classification, leveraging hard parameter sharing to improve the accuracy and robustness across MI tasks with minimal fine-tuning. Jin \textit{et al.}~\cite{jin2024cross} proposed a cross-dataset adaptive domain selection framework for MI-based BCIs, combining domain selection, data alignment, and enhanced common spatial patterns (CSP) to improve the classification accuracy while minimizing the calibration time. Liu \textit{et al.}~\cite{liu2025spatial} proposed SDDA, a framework based on spatial distillation and distribution alignment, specifically designed to address the heterogeneity and large EEG discrepancies.

All the methods discussed above, except for those proposed by Wu~\cite{drwuTNSRE2016} and Liu~\cite{liu2025spatial}, handle EEG heterogeneity simply by selecting overlapping channels shared across datasets. Although Liu \textit{et al.}~\cite{liu2025spatial} effectively addressed the fundamental challenge of EEG heterogeneity, their approach relies on access to target-domain (downstream) data for alignment in pre-adaptation scenarios.
 
\subsection{EEG Foundation Models}

Wang \textit{et al.}~\cite{wang2024cbramod} proposed CBraMod, a criss-cross transformer–based EEG foundation model with asymmetric conditional positional encoding, pre-trained via patch-based masked EEG reconstruction on over 27,000 hours of heterogeneous data. Chen et al.~\cite{wan2023eegformer} proposed EEGFormer, a vector-quantized Transformer pretrained on 1.7 TB of heterogeneous EEG data to learn transferable and interpretable representations for diverse downstream BCI tasks. Jiang \textit{et al.}~\cite{jiang2024large} proposed LaBraM, a large EEG foundation model that segments signals into channel patches, employs vector-quantized neural spectrum prediction for semantic tokenization, and leverages masked EEG modeling to pre-train Transformers on over 2,500 hours of diverse EEG data. Wang \textit{et al.}~\cite{wang2025eegpt} proposed EEGPT, a 10 million-parameter pretrained transformer that uses spatio-temporal representation alignment and mask-based reconstruction to learn universal EEG features. Jiang \textit{et al.}~\cite{jiang2024neurolm} proposed NeuroLM, a universal multi-task EEG foundation model that treats EEG signals as a foreign language via text-aligned neural tokenization.

Most of the above EEG foundation models integrate EEG data from multiple paradigms during training. However, substantial differences exist among EEG paradigms regarding data acquisition methods, active cortical regions, underlying neurophysiological principles, and relevant frequency bands. Additionally, EEG data quality varies significantly across datasets. Considering that the required EEG paradigm is typically known before downstream data acquisition, it is particularly important to design a paradigm-specific data construction pipeline to address these challenges effectively.

\section{CLEAN-MI}  \label{sect:method}

\begin{figure*}[htbp]\centering
\includegraphics[width=\linewidth,clip]{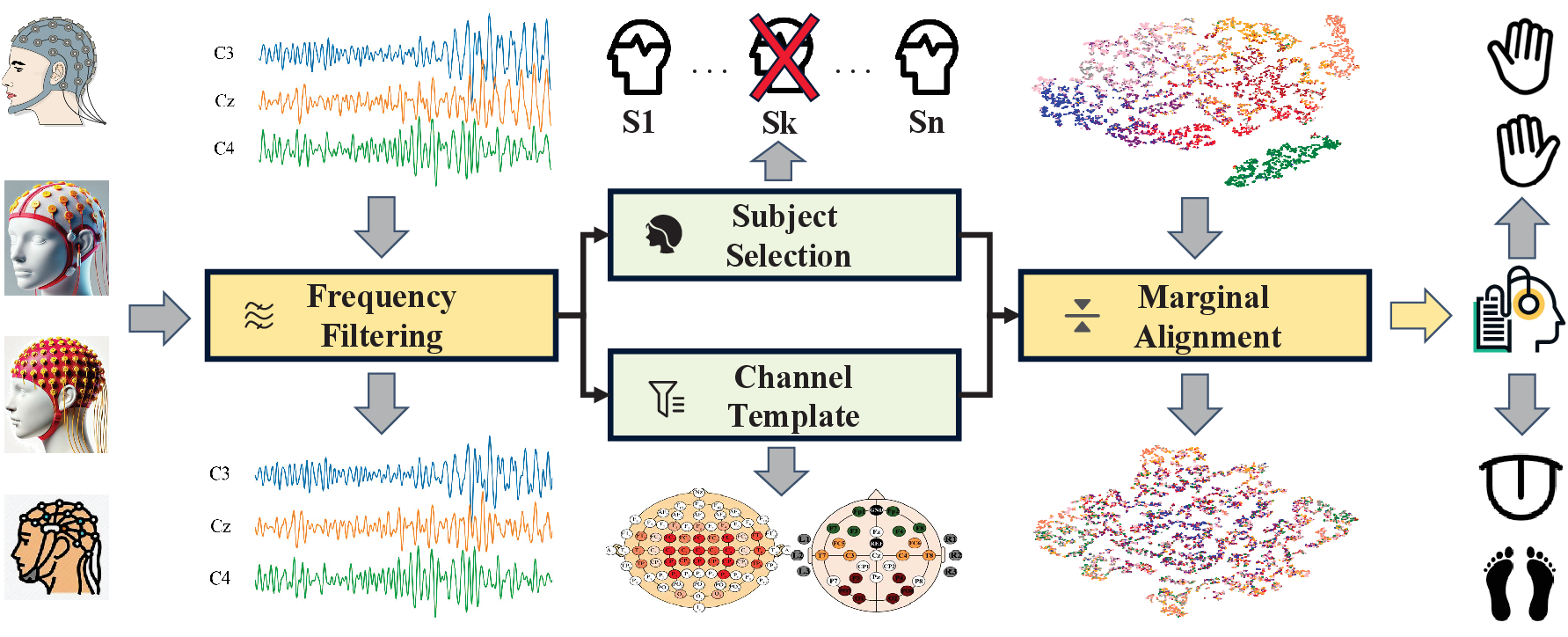}
\caption{Overview of the CLEAN-MI pipeline. EEG signals collected from various headsets are first filtered to retain motor-relevant frequency bands. Subject selection is then performed to remove low-quality subjects. Simultaneously, channel template alignment unifies electrode configurations across devices. Finally, marginal distribution alignment is applied to reduce domain shifts among subjects, yielding a consistent and discriminative feature space for motor imagery classification.} \label{fig:CLEAN-MI}
\end{figure*}

This section introduces our proposed CLEAN-MI for \textbf{C}onstructing \textbf{L}arge-scale, \textbf{E}fficient, and \textbf{A}ccurate \textbf{N}eurodata in the \textbf{MI} paradigm, as illustrated in Fig.~\ref{fig:CLEAN-MI}.

\subsection{EEG Data Collection}

\begin{figure}[htbp]         \centering
\includegraphics[width=\linewidth,clip]{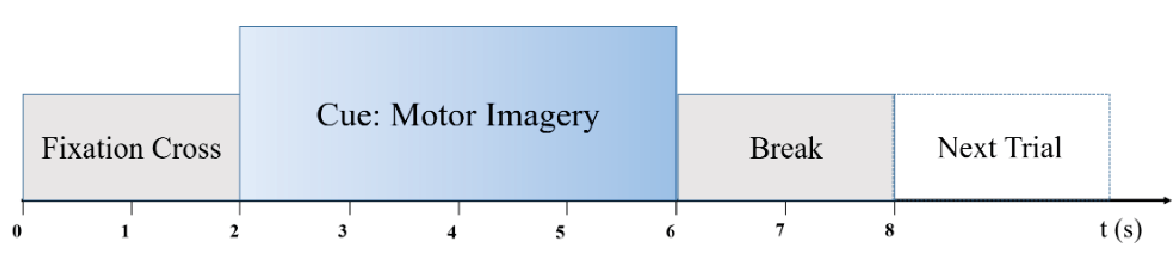}
\caption{MI paradigm data collection process.} \label{fig:data_acq}
\end{figure}

The data acquisition process for the MI paradigm involves the collection of EEG signals from the subjects, as illustrated in Fig.~\ref{fig:data_acq}. To acquire the EEG signals, the experimental setup includes the comfortable environment and the proper placement of the headset on the scalp of the subject. The subject is seated comfortably in a chair facing a computer screen. The subject is then instructed to perform a series of motor imagery tasks based on visual cues displayed on the screen. 

Each trial begins with the presentation of a fixation cross (`+'), signaling the subject to prepare for the upcoming MI task (t = 0). After a brief preparation period, an arrow appears on the screen pointing either left or right (other tasks may also be included such as feet and tongue). The direction of the arrow indicates the specific MI task to be performed. For instance, a rightward arrow prompts the subject to imagine right-hand movement, while a downward arrow corresponds to imagining foot movements (t = 2). The subject is expected to begin imagining the specified body part's movement immediately upon the arrow’s appearance and continue until the arrow disappears (t = 6). Following this, the fixation cross disappears, and the subject may rest briefly until the next trial begins (t = 8).

Notably, EEG headsets, sampling rates, and trial durations vary across datasets depending on the recording hardware and experimental protocol; therefore, harmonizing these parameters during the preprocessing stage is indispensable.

\subsection{Frequency Filtering}

The corresponding frequencies and their effects on behavior are summarized in Table ~\ref{tab:frequency_bands}. 

\begin{table}[htbp]
\centering
\caption{Frequency bands and their characteristics in MI paradigm}
\label{tab:frequency_bands}
\renewcommand{\arraystretch}{1.5} 
\begin{tabular}{c|c|l}
\toprule
\textbf{Band} & \textbf{Range (Hz)} & \textbf{Characteristics and Associated Regions} \\ \hline
$\delta$ & 0.5--4  & Deep sleep, unconscious states \\ \hline
$\theta$ & 4--8    & Relaxation, motor imagery, meditation \\ \hline
$\alpha$ & 8--13   & Sensorimotor areas, relaxation, motor imagery \\ \hline
$\beta$  & 13--30  & Central sensorimotor regions, motor control \\ \hline
$\gamma$ & 30--45  & Higher cognition, sensory processing \\ 
\bottomrule
\end{tabular}
\end{table}

Among these, the $\alpha$ (8–13 Hz) and $\beta$ (13–30 Hz) rhythms—originating from the sensorimotor cortices—are most strongly modulated by motor imagery. Kinesthetic imagery of movement induces event‐related desynchronization (ERD), i.e., a transient power decrease in these bands, whereas cessation of imagery elicits event‐related synchronization (ERS), i.e., a power rebound, typically with contralateral dominance in the $\beta$ band ~\cite{maeder2012pre}. Consequently, we employ an 8–30 Hz band‐pass filter to isolate these sensorimotor components.

\subsection{Channel Template}

The number and configuration of electrodes in EEG headsets vary across different models, often resulting in diverse channel setups. Each brain region is primarily responsible for controlling different behaviors. 

MI signals are associated with the phenomena of event-related desynchronization (ERD) and event-related synchronization (ERS). Specifically, when a subject imagines performing a movement, there is a decrease in the power of specific frequency bands (typically in the alpha and beta bands) in the brain regions associated with the imagined movement. This reduction in power is called event-related desynchronization (ERD) and is typically observed over the sensorimotor cortex, indicating a state of cortical activation. Conversely, if there is no movement imagery, certain brain areas may exhibit an increase in the power of these frequency bands, known as event-related synchronization (ERS). ERD is commonly observed during MI tasks, reflecting the mental preparation or intention to perform a motor action, whereas ERS may be associated with rest or a lack of motor activity ~\cite{blankertz2010neurophysiological}.

\begin{figure}[htbp]         \centering
\includegraphics[width=\linewidth,clip]{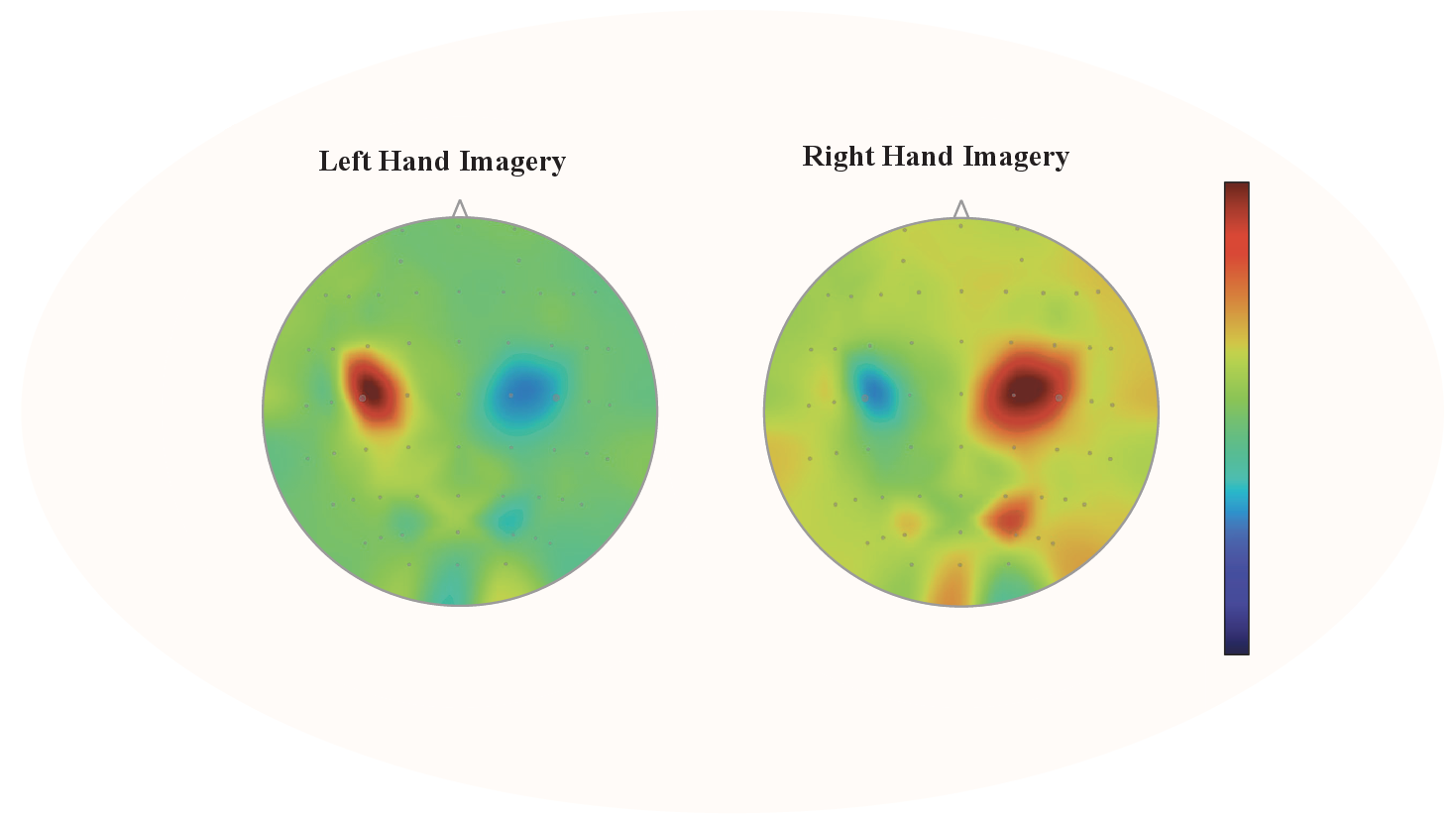}
\caption{Scalp topographies of SMR power changes during motor imagery of the left and right hands. The left panel shows spectral power decreases (blue) predominantly over the right hemisphere during left hand imagery, while the right panel shows power decreases (blue) over the left hemisphere during right hand imagery. The color bar indicates relative amplitude change in the SMR band, with blue denoting power attenuation and red denoting power increase.} \label{fig:SMR}
\end{figure}

MI tasks involving left-hand and right-hand movements typically show ERD over the C4 and C3 regions, respectively. Fig.~\ref{fig:SMR} depicts the phenomenon, ERD is observed in the right hemisphere during left-hand imagery and in the left hemisphere during right-hand imagery. These findings are fundamental to BCI systems that decode movement imagery signals from different limbs based on these cortical signatures. 

\begin{figure}[htbp]         \centering
\includegraphics[width=0.76\linewidth,clip]{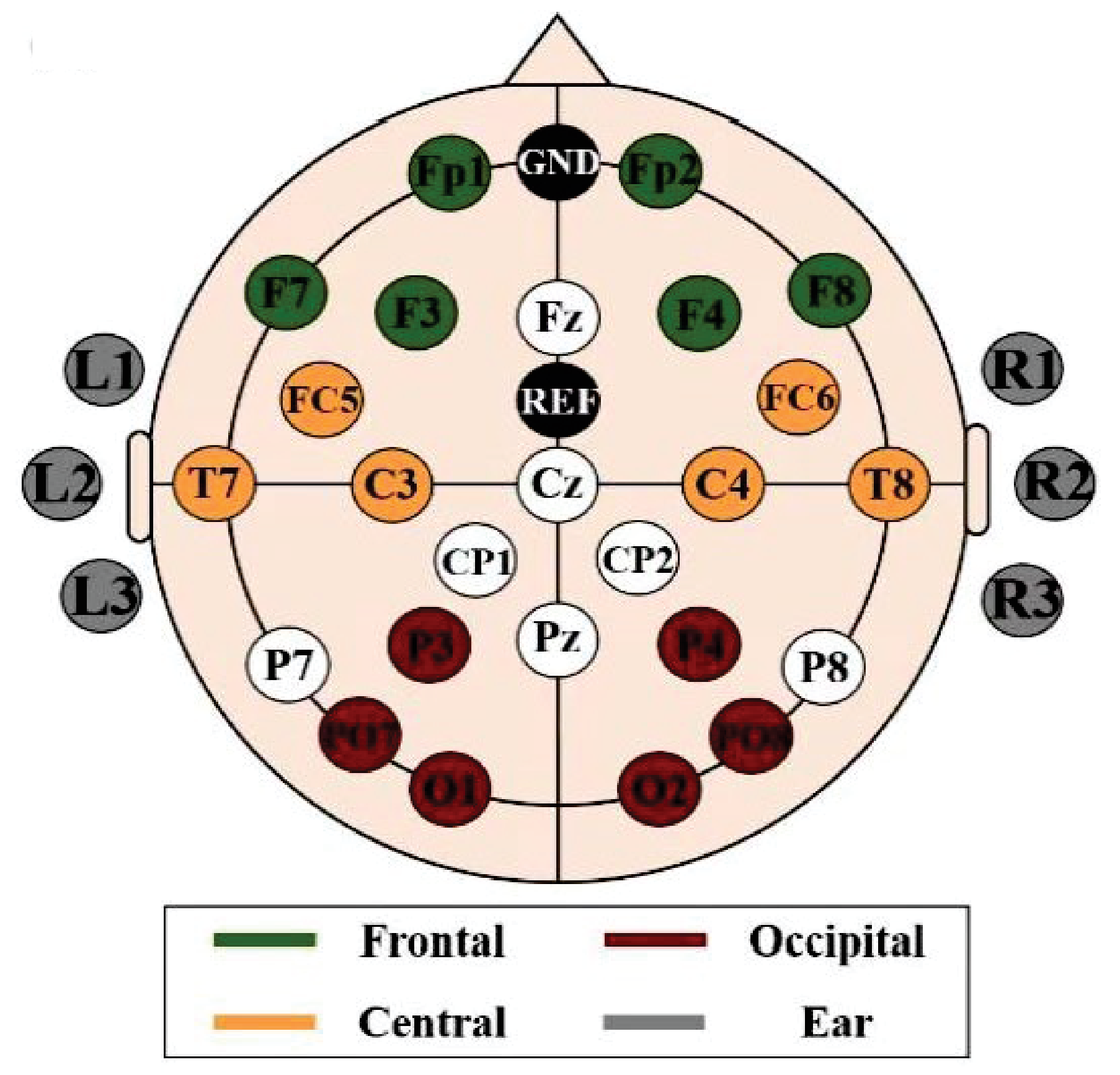}
\caption{Schematic of the EEG headset electrode positions.} \label{fig:channel_pos}
\end{figure}

Specifically, as indicated in Fig.~\ref{fig:channel_pos}, electrodes over the parietal (P) region are primarily engaged in visual processing and have been shown to contribute to EEG-based image reconstruction. Frontal (F) electrodes reflect attentional and executive functions, whereas central (C) electrodes directly overlie the sensorimotor cortex and are most informative for motor imagery tasks.

Due to non-invasive acquisition, MI EEG signals suffer from low signal-to-noise ratio and volume conduction, causing activity in the central motor cortex to spread to adjacent regions. To capture the most informative channels for MI decoding while mitigating spatial smearing, we define a template comprising electrodes over the frontal-central (FC), central (C), centro-parietal (CP), and temporal (T) regions. This selection emphasizes the sensorimotor cortex—where event-related desynchronization and synchronization are most pronounced—while excluding channels less relevant to MI, thereby reducing computational load and improving signal quality for downstream MI foundation model pre-training.

\subsection{Time Sample Alignment}

EEG datasets often differ in sampling rates and trial lengths, which hinders the generalization of methods that assume uniform temporal dimensions. While transfer learning within a single dataset can handle cross-subject or cross-session variability under a fixed sampling rate, cross-dataset scenarios introduce additional discrepancies in both sampling frequency and recording duration. To address this, all EEG recordings are resampled to a common rate (e.g., 200 Hz or 250 Hz) and trials are truncated or zero-padded to a fixed length~\cite{xie2023cross}. This temporal normalization harmonizes the time axis across diverse datasets, enabling seamless integration into foundation-model training pipelines.

\subsection{Expert Subject Selection}

EEG datasets often do not systematically evaluate the attention level and engagement of subjects prior to data acquisition, resulting in varying degrees of data quality. Additionally, differences in recording environments across laboratories or institutions further introduce variability and noise into EEG recordings. Inattention, fatigue, or distractions experienced by subjects, as well as environmental noise such as electrical interference or ambient sound, significantly degrade the quality and reliability of EEG signals. Therefore, we identify and select “expert subjects”, those whose EEG recordings are consistently high‐quality and informative to enhance overall data quality.

Specifically, we propose an expert subject selection procedure based on an initial within-subject evaluation experiment. For each participant, we train a classification model solely on their own EEG data collected during standard MI tasks. Subjects whose individual classification accuracies fall below a predefined threshold (typically set to 0.6) are excluded from further analysis. This selection criterion effectively identifies and removes subjects whose EEG recordings are substantially impaired by inattention, artifacts, or other adverse factors, resulting in a subset of reliable, high-quality expert subjects. Employing this strategy substantially reduces noise and enhances the robustness and interpretability of subsequent analyses, ultimately benefiting the development and performance of MI foundation models.

\subsection{Marginal Distribution Alignment}

EEG data are inherently non-stationary. Data normalization, often referred to as whitening, is a commonly employed preprocessing technique in machine learning to suppress noise. It not only helps mitigate marginal distribution shifts between the source and target domains, but also enhances the consistency within the source domain, particularly when EEG data are collected from multiple subjects.

Assume a subject has $n$ EEG trials $\{X_i\}_{i=1}^n$. EA first computes the mean covariance matrix of all trials:

\begin{align}
\bar{R}=\frac{1}{n}\sum_{i=1}^n X_iX_i^\top, \label{eq:EA-Ref}
\end{align}
and then performs the transformation:
\begin{align}
\widetilde{X}_i = \bar{R}^{-1/2} X_i. \label{eq:s-EA}
\end{align}

The mean covariance matrix of $\{\widetilde{X}_i\}_{i=1}^n$ becomes an identity matrix, i.e., the discrepancy in second-order statistics are reduced. $\{\widetilde{X}_i\}_{i=1}^n$ are then used to replace the original trials $\{X_i\}_{i=1}^n$ in all subsequent calculations.

\begin{figure*}[htbp]\centering
\subfigure[]{\includegraphics[width=0.46\linewidth,clip]{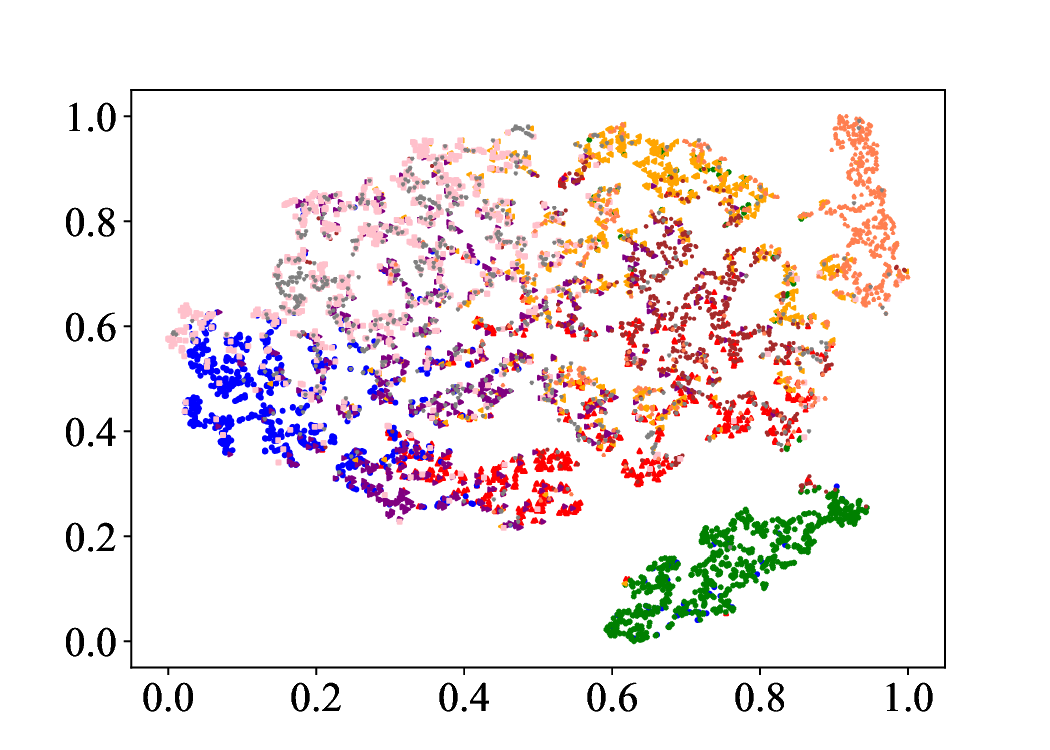}}
\subfigure[]{\includegraphics[width=0.52\linewidth,clip]{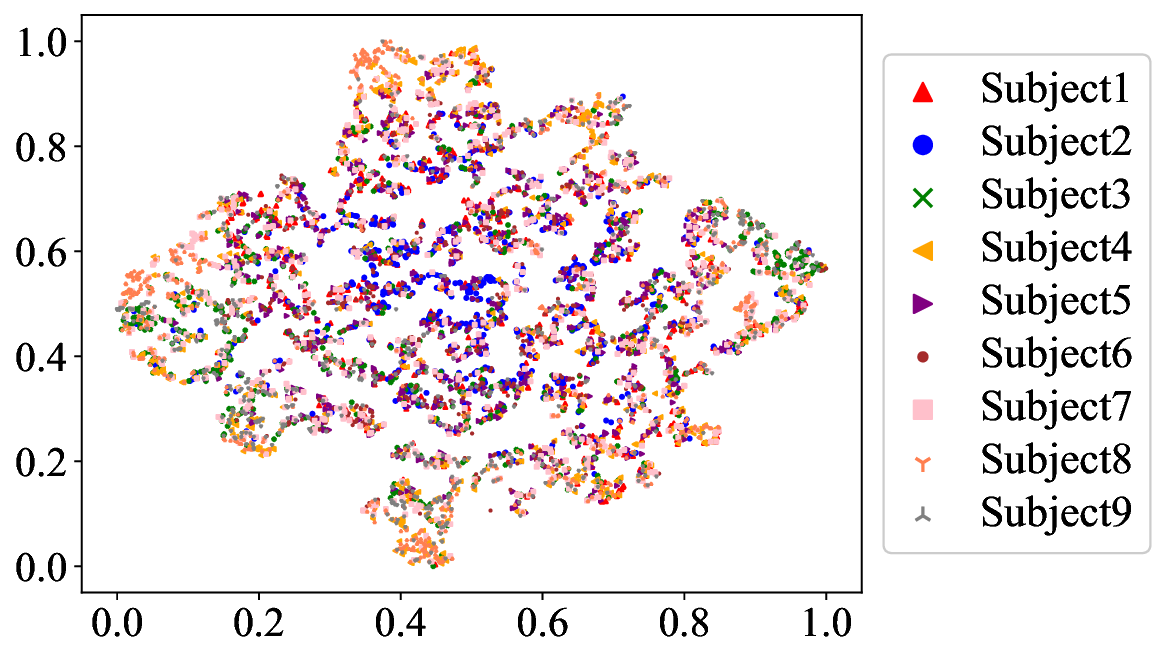}}
\caption{$t$-SNE visualization of the data in BNCI2014004. (a) Before EA; (b) After EA. Different colors represent trials from different subjects.} \label{fig:EA_tsne}
\end{figure*}

The rationale behind EA comprises two aspects: (1) EA alignment transforms the average covariance of each subject's trials into an identity matrix, where only the diagonal elements are non-zero. This transformation reduces the correlation between different channels and minimizes spatial redundancy, thus aiding the extraction of efficient feature representations. (2) EA alignment can be viewed as aligning each subject's information to a common point in the Riemannian space, which in the Euclidean space results in the transformed trials being evenly distributed across the same spatial distribution, as is shown in Fig.~\ref{fig:EA_tsne}.

\section{MI Datasets} \label{sect:Datasets}

This section introduces 18 MI benchmark datasets their SOTA approaches.

The datasets listed in Table ~\ref{tab:datasets} represent a diverse range of subjects, experimental setups, and MI tasks. These datasets have been widely used in the development and evaluation of algorithms, providing valuable insights and benchmarks for both traditional and deep learning approaches~\cite{jayaram2018moabb}.

\begin{enumerate}
\item  AlexMI~\cite{alexandre2006commande}: AlexMI dataset contains EEG recordings from 8 subjects, performing 2 task of motor imagination (right hand, feet or rest). Data have been recorded at 512Hz with 16 wet electrodes (Fpz, F7, F3, Fz, F4, F8, T7, C3, Cz, C4, T8, P7, P3, Pz, P4, P8).
\item BNCI2014001~\cite{tangermann2012review}: BNCI2014001 dataset contains EEG data from 9 subjects performing four MI tasks: left hand, right hand, both feet, and tongue. Each subject participated in two sessions, with each session consisting of 6 runs, yielding a total of 288 trials per session. The SOTA algorithm  is available on \footnote{https://paperswithcode.com/sota/within-session-motor-imagery-all-classes-on-2}.
\item BNCI2014004~\cite{leeb2007brain}: This dataset includes EEG data from 9 right-handed subjects, who performed two MI tasks: left hand and right hand. Each subject participated in five sessions, with the first two for screening without feedback and the last three with feedback. The data was recorded with three bipolar EEG channels (C3, Cz, C4) at 250 Hz and included 120 trials per subject for each MI category. The SOTA algorithm  is available on \footnote{https://paperswithcode.com/sota/within-session-motor-imagery-left-hand-vs-1}.
\item BNCI2014002~\cite{steyrl2016random}: BNCI2014002 dataset includes EEG data from 13 participants performing sustained MI of the right hand and feet. The session consists of eight runs, with 50 trials per class for training and 30 trials for validation. EEG was recorded at 512 Hz from 15 electrodes, including C3, Cz, and C4, with a biosignal amplifier and active Ag/AgCl electrodes. The SOTA algorithm  is available on \footnote{https://paperswithcode.com/dataset/bnci2014-002-moabb-1}.
\item BNCI2015001~\cite{faller2012autocalibration}: This dataset contains EEG data from subjects performing sustained MI of the right hand and both feet. The data were recorded at 512 Hz using 15 electrodes, including C3, Cz, and C4, with a bandpass filter between 0.5 and 100 Hz and a notch filter at 50 Hz. The SOTA algorithm  is available on \footnote{https://paperswithcode.com/sota/within-session-motor-imagery-right-hand-vs-3}.
\item BNCI2015004~\cite{scherer2015individually}: BNCI2015004 dataset includes EEG data from 9 users with disabilities performing five mental tasks: word association, subtraction, spatial navigation, and motor imagery of the right hand and feet. Data were recorded at 256 Hz from 30 electrodes, with a 0.5-100 Hz bandpass filter and a 50 Hz notch filter. The SOTA algorithm  is available on \footnote{https://paperswithcode.com/sota/within-session-motor-imagery-right-hand-vs-4}.
\item Cho2017~\cite{cho2017eeg}: Cho2017 dataset includes EEG data from 52 subjects (19 females, mean age 24.8 ± 3.86 years) performing MI tasks for the left and right hands. EEG was recorded at 512 Hz from 64 channels using the Biosemi ActiveTwo system, with a 10-10 system montage.
\item Lee2019~\cite{lee2019eeg}: Lee2019 dataset includes EEG data recorded from 62 channels at 1,000 Hz using a BrainAmp amplifier, which involved MI tasks for left and right hand grasping, with 100 trials per session. The EEG channels were referenced to the nasion and grounded to AFz.
\item GrosseWentrup2009~\cite{grosse2009beamforming}: GrosseWentrup2009 dataset includes EEG data from 10 healthy subjects (8 right-handed, mean age 25.6 ± 2.5 years) performing haptic MI tasks for the left and right hands. EEG was recorded at 500 Hz from 128 electrodes placed according to the extended 10-20 system, with Cz as the reference.
\item Ofner2017 ~\cite{ofner2017upper}: Ofner2017 dataset includes EEG data from 15 healthy subjects (mean age 27 ± 5 years) performing motor execution (ME) and motor imagery (MI) tasks. Subjects performed six movement types with the right upper limb, including elbow flexion/extension, forearm supination/pronation, and hand open/close, across two sessions recorded on different days. The dataset also includes a rest condition where no movement was performed.
\item PhysionetMI ~\cite{goldberger2000physiobank}: PhysionetMI dataset includes over 1500 one- and two-minute EEG recordings from 109 volunteers performing MI tasks. EEG was recorded with 64 channels using the BCI2000 system~\cite{schalk2004bci2000}.
\item Schirrmeister2017 ~\cite{schirrmeister2017deep}: Schirrmeister2017 dataset includes EEG data from 14 healthy subjects (mean age 27.2 ± 3.6 years), recorded using 128 electrodes, of which 44 electrodes covering the motor cortex were used for analysis. Subjects performed four types of movements (left hand, right hand, both feet, and rest) in approximately 1000 four-second trials, divided into 13 runs per subject.
\item Shin2017A ~\cite{shin2016open}: Shin2017A dataset includes EEG and NIRS data collected from 30 subjects using a BrainAmp EEG amplifier at 1000 Hz sampling rate, with electrodes placed according to the 10-5 system.
\item Shin2017B ~\cite{shin2016open}: Same as Shin2017A dataset.
\item Weibo2014 ~\cite{yi2014evaluation}: Weibo2014 dataset includes EEG data from 10 subjects recorded with 60 electrodes. It consists of seven mental tasks, including simple and compound limb MI tasks (left hand, right hand, feet, and combinations), and a rest state. The SOTA algorithm  is available on \footnote{https://paperswithcode.com/dataset/weibo2014-moabb}.
\item Zhou2016 ~\cite{zhou2016fully}: Zhou2016 dataset includes EEG data from 4 subjects performing three MI tasks: left hand, right hand, and feet. Each subject participated in three sessions, with each session consisting of two runs of 75 trials (25 trials per class). The SOTA algorithm  is available on \footnote{https://paperswithcode.com/dataset/zhou2016-moabb}.
\item Stieger2021 ~\cite{stieger2021continuous}: Stieger2021 dataset includes EEG data from 62 participants (33 MBSR participants and 29 controls) who underwent MI-based BCI training, following an 8-week mindfulness intervention or a waitlist control condition. The dataset focuses on how individuals learn to control SMR-BCIs, with participants completing 6 to 10 sessions of BCI training after the intervention.
\item Liu2024 ~\cite{liu2024eeg}: Liu2024 dataset includes EEG data from 50 acute stroke patients (mean age 56.7 ± 10.57 years), recorded during a MI experiment with left and right hand movements. EEG was collected using a wireless 29-electrode system at 500 Hz, with trials consisting of 8-second tasks alternating between instruction, motor imagery, and break stages.
\end{enumerate}

\begin{table*}[htpb]
\centering
\caption{Summary of the MI benchmark datasets.}
\label{tab:datasets}
\begin{tabular}{c|c|c|c|c|c|c|c}
\toprule
\multirow{2}{*}{Dataset} & Number of & Number of & Sampling & Trial Length & Number of Trials & \multirow{2}{*}{Paradigm} & \multirow{2}{*}{Classes} \\
 & Subjects & Channels & Rate (Hz) & (seconds) & in a Session &  &  \\
 \midrule
AlexMI & 8 & 16 & 512 & 3  & 20 & MI & 3 \\
\midrule
BNCI2014001 & 9 & 22 & 250 & 4  & 144 & MI & 2 \\
\midrule
BNCI2014004 & 9 & 3 & 250 & 4 & 680-760 & MI & 2 \\
\midrule
BNCI2014002 & 14 & 15 & 512 & 5 & 100 & MI & 2 \\
\midrule
BNCI2015001 & 12 & 13 & 512 & 5 & 200 & MI & 2 \\
\midrule
BNCI2015004 & 9 & 30 & 256 & 7 & 80 & MI & 5 \\
\midrule
Cho2017 & 52 & 64 & 512 & 3 & 200-240 & MI & 2 \\
\midrule
Lee2019 & 54 & 62 & 1000 & 4 & 100 & MI & 2 \\
\midrule
GrosseWentrup2009 & 10 & 128 & 500 & 7 & 150 & MI & 2 \\
\midrule
Ofner2017 & 15 & 61 & 512 & 3 & 60 & MI & 7 \\
\midrule
PhysionetMI & 109 & 64 & 160 & 3 & 23 & MI & 4 \\
\midrule
Schirrmeister2017 & 14 & 128 & 500 & 4 & 120 & MI & 4 \\
\midrule
Shin2017A & 29 & 30 & 200 & 10 & 30 & MI & 2 \\
\midrule
Shin2017B & 29 & 30 & 200 & 10 & 30 & MI & 2 \\
\midrule
Weibo2014 & 10 & 60 & 200 & 4 & 80 & MI & 7 \\
\midrule
Zhou2016 & 4 & 14 & 250 & 5 & 160 & MI & 3 \\
\midrule
Stieger2021 & 62 & 64 & 1000 & 3 & 450 & MI & 4 \\
\midrule
Liu2024 & 50 & 29 & 500 & 4 & 20 & MI & 2 \\
\bottomrule
\end{tabular}
\end{table*}

\section{Experiments} \label{sect:exp}
\subsection{Experimental Settings}

To evaluate the effectiveness of the proposed CLEAN-MI pipeline, we conducted experiments on three public motor imagery datasets: Weibo2014, Cho2017, and BNCI2015001. For all datasets, EEG signals were first bandpass filtered in the 8--30~Hz range to isolate the $\alpha$ and $\beta$ rhythms, which are known to be most relevant for motor imagery tasks.

\textbf{Channel Template.} We utilized MI-relevant channels based on spatial neurophysiological priors described in Section~\ref{sect:method}. 

\begin{itemize}
    \item In the \textit{Weibo2014} dataset (60 channels), we selected the following 35 channels: FT7, FC5, FC3, FC1, FCZ, FC2, FC4, FC6, FT8, T7, C5, C3, C1, Cz, C2, C4, C6, T8, TP7, CP5, CP3, CP1, CPz, CP2, CP4, CP6, TP8, P7, P5, P3, P1, Pz, P2, P4, P6, P8.
    \item In the \textit{Cho2017} dataset (64 channels), we selected 38 MI-related channels: FT7, FC5, FC3, FC1, FCZ, FC2, FC4, FC6, FT8, T7, C5, C3, C1, Cz, C2, C4, C6, T8, TP7, CP5, CP3, CP1, CPz, CP2, CP4, CP6, TP8, P9, P7, P5, P3, P1, Pz, P2, P4, P6, P8, P10.
    \item For the \textit{BNCI2015001} dataset, which contains only 13 channels, all channels fall within MI-relevant cortical regions. Thus, no additional channel filtering was applied.
\end{itemize}

\textbf{Subject Selection.} To remove noisy or low-quality subjects that may negatively impact model performance, we applied within-subject validation. Each subject’s data were randomly split into training and testing sets with an 8:2 ratio. Subjects with classification accuracies below threshold (typically set to 0.6) were excluded from further training.

\begin{itemize}
    \item In the \textit{Weibo2014} dataset, we excluded subjects S2, S3, S4, and S9.
    \item In the \textit{Cho2017} dataset, we excluded subjects S1, S6, S7, S12, S15, S16, S26, S27, S28, S31, S32, S33, S34, S36, S38, S39, and S48.
    \item In the \textit{BNCI2015001} dataset, we excluded subject S7.
\end{itemize}

\textbf{Model and training settings.} We adopted EEGNet~\cite{lawhern2018eegnet} as the backbone model for all experiments. Hyperparameters are consistent across datasets: batch size was set to 32, learning rate to 0.001, and number of training epochs to 100. During subject screening, within-subject validation was applied as described above. For final performance evaluation, we used a leave-one-subject-out (LOSO) cross-validation strategy, where each subject was iteratively held out for testing while the remaining subjects were used for training. This protocol simulates a realistic cross-subject adaptation scenario and demonstrates the generalizability of the CLEAN-MI constructed data.

\subsection{Results}

Tables~\ref{tab:MI_1}--\ref{tab:MI_3} present the experimental results across multiple MI datasets. The proposed CLEAN-MI pipeline demonstrates clear advantages in both computational efficiency and classification performance. By leveraging the channel template and expert subject selection, our approach not only reduces computational cost but also improves model accuracy. For example, on the Cho2017 dataset, the computational complexity was reduced by 50\%--70\%, while the classification accuracy improved by 1.5 percentage points. This simultaneous reduction in computational overhead and enhancement of model performance is particularly encouraging for large-scale foundation model research.

\begin{table*}[htpb]     \centering
\fontsize{8}{10}\selectfont %
       \caption{Classification accuracies (\%) using raw data and subject selection on BNCI2015001. The best accuracies are marked in bold.}  \label{tab:MI_1}
    \setlength{\tabcolsep}{3.5pt} 
    \begin{tabular}{c|ccccccccccccc}   \toprule
        Setting & S0 & S1 & S2 & S3 & S4 & S5 & S6 & S7 & S8 & S9 & S10 & S11 & Avg. \\
        \midrule
         \multirow{1}{*}{Raw Data}
         & 92.33 & \textbf{96.83} & 67.5 & \textbf{86.33} & \textbf{90.67} & 66.50 & 72.83 & 65.00 & \textbf{65.00} & 67.83 & \textbf{62.50} & 51.50 & 73.74$_{\pm0.73} (74.53)$ \\
        \midrule
        \multirow{1}{*}{Channel Template}
        & \textbf{93.00} & 95.67 & \textbf{80.67} & 86 & 89.5 & \textbf{71.83} & 72.5 & --- & 63.5 & \textbf{69.67} & 56.50  & \textbf{53.67} &  \textbf{75.68}$_{\pm1.32}$\\
        \bottomrule
    \end{tabular}
\end{table*}

\begin{table*}[htpb]     \centering
\fontsize{8}{10}\selectfont %
       \caption{Classification accuracies (\%) using raw data and CLEAN-MI processing steps on Weibo2014. The best accuracies are marked in bold.}  \label{tab:MI_2}
    \setlength{\tabcolsep}{3.5pt} 
    \begin{tabular}{c|ccccccccccc}   \toprule
        Setting & S0 & S1 & S2 & S3 & S4 & S5 & S6 & S7 & S8 & S9 & Avg. \\
        \midrule
         \multirow{1}{*}{Raw Data}
         & 66.25 & 75.13 & 52.00 & 51.50 & 53.38 & 93.00 & 84.13 & 56.88 & 73.13 & 52.63 & 65.8$_{\pm1.01} (74.75)$ \\

        \midrule
        \multirow{1}{*}{Channel Template}
        & 64.13 & 80.00 & 55.50 & 49.38 & 51.50 & 92.71 & 88.38 & 77.38 & \textbf{77.88} & 55.88 & 69.27$^{\text{*}}_{\pm0.74} (80.08)$\\

        \midrule
        \multirow{1}{*}{Subject Selection}
        & \textbf{67.50} & 80.63 & --- & --- & --- & 95.71 & 87.5 & 56.25 & 76.88 & --- & 77.41$^{\text{*}}_{\pm0.65}$\\

        \midrule
        \multirow{1}{*}{Subject Selection + Channel Template}
        & 66.25 & \textbf{84.38} & --- & --- & --- & \textbf{96.43} & \textbf{90.62} & \textbf{78.12} & 75.62 & --- &  \textbf{81.90}$^{\text{**}}_{\pm0.91}$\\

        \bottomrule
    \end{tabular}
    
    \vspace{1mm}  
    \parbox{0.95\textwidth}{
        \scriptsize
        \textbf{Note:} ****: $p < 0.0001$; ***: $p < 0.001$; **: $p < 0.01$; *: $p < 0.05$.
    }   
\end{table*}

\begin{table*}[htpb]
\centering
\fontsize{8}{10}\selectfont
\caption{Classification accuracies (\%) using raw data and CLEAN-MI processing steps on Cho2017 dataset. The best accuracies are marked in bold.}
\label{tab:MI_3}
\setlength{\tabcolsep}{3.5pt} 
\begin{tabular}{c|cccccccccccccc}
\toprule
\textbf{Setting} & S0 & S1 & S2 & S3 & S4 & S5 & S6 & S7 & S8 & S9 & S10 & S11 & S12 & S13 \\
\midrule
Raw Data & 64.60 & 55.10  & 92.70 & 87.90 & 66.30 & \textbf{62.70} & 57.92 & 58.10 & 76.00 & 85.50 & 63.80 & 62.80 & 52.80 & 86.30  \\
Channel Template & \textbf{66.4} & 56.7 & 92.40 & 90.10 & \textbf{70.00} & 60.50 & 60.17 & 58.90 & \textbf{76.92} & \textbf{86.90} & 66.40 & 64.90 & 48.30 & 85.80 \\
Subject Selection & 61.50 & --- & 92.10 & 89.80 & 67.00 & 58.30 & --- & --- & 76.75 & 86.80 & 65.70 & 65.70 & --- & \textbf{91.50} \\
Channel Template + Subject Selection & 65.60 & --- & \textbf{92.90} & \textbf{90.90} & 65.80 & 59.40 & --- & --- & 75.75 & 86.50 & \textbf{69.40} & \textbf{69.30} & --- & 90.50 \\
\bottomrule
\toprule
\textbf{Setting} & S14 & S15 & S16 & S17 & S18 & S19 & S20 & S21 & S22 & S23 & S24 & S25 & S26 & S27 \\
\midrule
Raw Data & 76.80 & 59.70  & 49.50 & 67.60 & 64.30 & 65.30 & 71.20 & 73.00 & 87.00 & 74.70 & 76.00 & 74.00 & 50.60 & 50.10  \\
Channel Template & 77.50 & 55.80 & 49.10 & 69.90 & 63.10 & \textbf{66.50} & \textbf{76.00} & 72.00 & \textbf{90.00} & 74.80 & 75.50 & 78.20 & 52.30 & 50.70 \\
Subject Selection & 78.20 & --- & --- & 69.50 & \textbf{66.00} & 66.40 & 70.50 & 72.50 & 88.70 & \textbf{74.90} & 77.80 & 77.90 & --- & --- \\
Channel Template + Subject Selection & \textbf{79.20} & --- & --- & \textbf{73.00} & 63.70 & 64.70 & 72.70 & \textbf{73.60} & 89.40 & 74.60 & \textbf{80.70} & \textbf{78.60} & --- & --- \\
\bottomrule
\toprule
\textbf{Setting} & S28 & S29 & S30 & S31 & S32 & S33 & S34 & S35 & S36 & S37 & S38 & S39 & S40 & S41 \\
\midrule
Raw Data & 53.50 & 62.50 & 66.90 & 51.90 & 59.50 & 54.50 & 57.30 & 65.50 & 50.80 & 62.00 & 56.70 & 50.90 & 91.70 & 71.50 \\
Channel Template & 53.20 & 63.60 & 66.30 & 50.60 & 64.60 & 55.40 & 56.60 & 65.40 & 49.20 & 61.30 & 58.50 & 48.50 & 93.80 & 72.20 \\
Subject Selection & --- & 63.20 & \textbf{69.70} & --- & --- & --- & --- & 65.50 & --- & \textbf{62.70} & --- & --- & 92.80 & 71.50 \\
Channel Template + Subject Selection & --- & \textbf{64.20} & 66.90 & --- & --- & --- & --- & \textbf{66.90} & --- & 61.00 & --- & --- & \textbf{94.50} & \textbf{72.70} \\
\bottomrule
\toprule
\textbf{Setting} & S42 & S43 & S44 & S45 & S46 & S47 & S48 & S49 & S50 & S51 & \multicolumn{3}{c}{Avg.} \\
\cmidrule(lr){1-14}
Raw Data & \textbf{96.90} & 71.00 & 67.30 & 69.42 & 66.10 & 94.90 & 63.70 & 59.80 & 60.90 & 65.70 & \multicolumn{3}{c}{$66.99_{\pm0.91} (72.99)$} \\
Channel Template & 96.30 & 73.10 & 66.80 & 72.25 & \textbf{68.90} & 94.60 & 65.40 & 59.40 & \textbf{61.30} & \textbf{67.40} & \multicolumn{3}{c}{$67.70_{\pm0.26} (74.07)$} \\
Subject Selection & 96.80 & 68.90 & \textbf{68.20} & 70.25 & 68.40 & 95.30 & 61.20 & --- & 58.40 & 66.90 & \multicolumn{3}{c}{$73.64^{\text{*}}_{\pm0.36}$} \\
Channel Template + Subject Selection & 96.60 & \textbf{73.60} & 67.20 & \textbf{72.33} & 68.30 & \textbf{96.00} & \textbf{65.60} & --- & 60.30 & 66.80 & \multicolumn{3}{c}{$\textbf{74.55}^{\text{*}}_{\pm0.39}$} \\
\cmidrule(lr){1-14}
\end{tabular}

\vspace{1mm}  
\parbox{0.95\textwidth}{
    \scriptsize
    \textbf{Note:} ****: $p < 0.0001$; ***: $p < 0.001$; **: $p < 0.01$; *: $p < 0.1$.
} 
\end{table*}

\section{Future Research Directions} \label{sect:future}

\subsection{Heterogeneous Euclidean Alignment} 

EA has proven to be highly effective for aligning EEG signals within a single dataset, particularly for reducing inter-subject variability. By transforming the data from different subjects into a common spatial distribution, EA significantly improves the consistency of feature extraction and classification performance. However, in the scenario of transfer learning and multi-task learning, a more challenging problem arises when attempting to align data from different datasets. Each dataset may be recorded using different EEG acquisition systems, with variations in electrode configurations, electrode numbers and positions. These differences introduce heterogeneous feature spaces, making it difficult to directly apply previous EA methods.

To address this issue, future research needs to focus on developing alignment techniques that can handle these heterogeneous feature spaces across datasets with different EEG setups. The goal is to map data from diverse sources into a shared distribution space while preserving the unique characteristics of each dataset. This problem is particularly critical when working with cross-dataset transfer learning, where the model needs to generalize across datasets with varying acquisition protocols. Solving this challenge will enable more robust and scalable BCI systems that can effectively use data from multiple sources without being biased by the specificities of individual datasets. Research in this area could lead to novel techniques for domain adaptation and alignment, allowing for better integration of EEG data from heterogeneous environments.

\subsection{Construct High-Quality MI Datasets}

The foundation of MI foundation model depends on the availability of large-scale, high-quality datasets. Training on clean, well-processed EEG data is critical for advancing model generalization, robustness, and transferability. However, EEG signals are inherently susceptible to noise, exhibit considerable inter-subject variability, and present significant heterogeneity in both spatial configuration and signal quality across different datasets.

To address these challenges, we propose CLEAN-MI, a scalable and efficient data construction pipeline specifically designed to extract high-quality MI-related EEG signals. By incorporating channel template selection and subject-level screening, CLEAN-MI systematically filters out irrelevant or low-quality data, preserving only the most informative and task-relevant EEG components.

In this study, we validated the effectiveness of CLEAN-MI on three representative MI datasets, demonstrating consistent improvements in data quality and model performance. In future work, we plan to extend the application of CLEAN-MI across a broader range of both public and proprietary MI datasets. Our goal is to construct the largest unified repository of high-quality MI EEG data to date, serving as a solid foundation for pretraining general-purpose MI foundation models. We also intend to release this high-quality dataset to the research community to facilitate further advancements in MI-based LM research.

\subsection{Foundation Model for MI Paradigms} 

Inspired by the remarkable success of large-scale models such as GPT~\cite{achiam2023gpt}, LLaMA~\cite{dubey2024llama}, and Qwen~\cite{bai2023qwen} in artificial intelligence, there has been increasing interest in developing foundation models specifically designed for brain–computer interfaces (BCIs). Recent models such as LaBraM~\cite{jiang2024large} and EEGPT~\cite{wang2025eegpt} represent early efforts to bring this paradigm to motor imagery (MI)-based BCIs. Despite these advances, current MI foundation models still face challenges related to generalization and robustness across diverse datasets.

To address these limitations, our future work will leverage the large-scale, high-quality MI dataset constructed via CLEAN-MI to pretrain MI foundation models with improved generalization capabilities. By training on standardized, high-quality data, we aim to establish a universal MI foundation model that is better suited to diverse downstream BCI tasks and real-world scenarios.

\section{Conclusions} \label{sect:conclusion}

In this paper, we propose CLEAN-MI, a scalable and systematic data construction pipeline for motor imagery (MI) EEG foundation models. Our approach integrates frequency band filtering, channel template selection, subject selection, and marginal distribution alignment to address the challenges of noise, heterogeneity, and variable data quality in multi-source MI EEG datasets. Experimental results on several public datasets demonstrate that CLEAN-MI consistently improves data quality and model performance. The proposed pipeline provides a robust foundation for developing generalizable MI foundation models and provides a practical framework for constructing large-scale, high-quality EEG datasets for future BCI research.

\end{document}